\documentclass[12pt]{article}
\usepackage{amsmath,slett,epsfig,macros,cite}

\newcommand{\half}[0]{\frac{1}{2}} 
\def\mq{m_{\rm q}}

\begin{document}

\begin{titlepage}

\begin{flushright}
HU-EP-05/67\\
SFB/CPP-05-71
\end{flushright}

\vskip 1.2cm
\begin{center}
{\Large\bf 
Cutoff effects of Wilson fermions in the absence of spontaneous chiral 
symmetry breaking \\[0.5ex]
}
\end{center}
\vskip 1.7cm
\begin{center}
{ \large
Michele Della Morte$^{\scriptscriptstyle a}$ and
Magdalena Luz$^{\scriptscriptstyle a}$
}
\vskip 0.5cm
{
$^{\scriptstyle a}$ 
Institut f\"ur Physik, Humboldt Universit\"at, \\
Newtonstr. 15, 12489 Berlin, Germany
\vskip 1.0ex
}
\vskip 1.575cm
{\bf Abstract}
\vskip 0.1ex
\end{center}
{\small
We simulate two dimensional QED with two degenerate Wilson fermions and 
plaquette gauge action.
As a consequence of the  Mermin-Wagner theorem, in the continuum limit chiral
symmetry is realized \`a la Wigner. This property affects also the
size of the cutoff effects. That can be understood in view of the fact
that the leading lattice artifacts are described, in the continuum Symanzik 
effective theory, by chirality breaking terms. In particular, 
vacuum expectation values of non-chirality-breaking operators are expected to
be O($a$) improved in the chiral limit. We provide a numerical confirmation
of this expectation by performing a scaling test.
}


\vskip 0.29cm
\vfill

\begin{center}
October 2005
\end{center}

\eject
\vfill
\eject

\end{titlepage}

\newcommand{\ssect}[1]{\noindent {\bf  #1.}}

\ssect{1}
Cutoff effects in lattice (gauge) theories can be described using an 
effective continuum action, as proposed by Symanzik in refs.~\cite{Symanzik_b,Sym_pap}.
In this approach the leading lattice artifacts (e.g. in the spectrum of the theory)
can be removed by including a set of irrelevant operators in the action and by 
properly tuning their coefficients. For the case of the Wilson lattice regularization
of QCD~\cite{Wilson}, the relevant coefficient can be tuned by requiring the 
restoration of chiral
 symmetry up to O($a^2$). This interplay between chiral symmetry and cutoff effects 
has been addressed in detail in ref.~\cite{Luscher:1996sc}.

Further insights on this connection have been recently derived in 
ref.~\cite{Frezzotti:2003ni} by considering also so called spurionic lattice symmetries
to classify the operators, which can appear in the Symanzik effective theory.
Without reproducing the whole argument, here we will 
simply summarize the results relevant as premises for this work. 
Let's consider  the Wilson fermionic action for two degenerate flavors
\begin{eqnarray}
&& S_{\rm F}[U,\psibar,\psi]=a^4 \sum_x \psibar(x)(D+m_0)\psi(x) \quad{\rm and} \; \\
&&D=\half\left\{ \gamma_{\mu}(\nabla^*_{\mu}+\nabla_{\mu})-ar\nabla^*_{\mu}\nabla_{\mu}
\right\} \,, 
\end{eqnarray}
where $\nabla^*_{\mu}$ and $\nabla_{\mu}$ are the covariant backward and forward 
lattice derivative respectively, $U$ denotes the gauge field  
 and $r$ is the Wilson parameter, which we set to 1.
The vacuum expectation value of a multiplicatively
renormalizable operator $\cal{O}$ can be expanded as
\be
\langle \mathcal{O} \rangle |_{r,m_{\rm q}} =\left[\zeta^\mathcal{O}+ a\mq \xi^\mathcal{O}
\right] \langle \mathcal{O} \rangle |^{\rm cont}_{\mq}+a \sum \limits_l
(\mq)^{n_l} \eta^{\mathcal{O}}_{\mathcal{O}_l} \langle \mathcal{O}_l 
\rangle |^{\rm cont}_{\mq} +{\rm O}(a^2)\;,    
\label{Sym}
\ee
where  $\mq$ is the bare subtracted fermion mass defined as
$\mq=m_0-m_{\rm c}$, such that the physical fermion mass vanishes for $m_0=m_{\rm c}$.
We refer to~\cite{Frezzotti:2003ni} for any unexplained notation in eq.~(\ref{Sym}).
The operators $\mathcal{O}_l$ appearing on the rhs result from the  insertion of the 
O($a$) terms in the action and from the O($a$) terms associated with
the operator $\mathcal{O}$ itself.
They can be classified according to their parity $P_{R_5}^{\mathcal{O}_l}$ under the $R_5$ 
transformation
\be
R_5\;:\; \psi \rightarrow \psi'=\gamma_5 \psi \; , \qquad \psibar \rightarrow 
\psibar'=-\psibar \gamma_5 \;,
\ee
which is a non-anomalous element of the chiral group and produces a spurionic symmetry of the 
Wilson action when combined with the replacements $r\rightarrow -r$, $\mq \rightarrow -\mq$. 
The authors  of ref.~\cite{Frezzotti:2003ni} have shown that
\be
P_{R_5}^{\mathcal{O}}+P_{R_5}^{\mathcal{O}_l}+n_l = 1 \; {\rm mod}\; (2),
\label{parities}
\ee
which, loosely speaking, implies that the O($a$) terms in the chiral limit (where the sum 
reduces to the $n_l=0$ contributions) have  
opposite $R_5$-parity compared to the leading term.
It is indeed interesting to consider the limit $\mq \to 0$ in eq.~(\ref{Sym}). Two different 
scenarios are possible
\begin{itemize}
\item
Spontaneous breaking of chiral symmetry  does not occur, 
as for QCD in small volume, therefore the theory is analytical at $\mq=0$ and
we can directly set the fermion mass to zero in eq.~(\ref{Sym}).  From that we infer
that if $\mathcal{O}$ is even under $R_5$ then the operators $\mathcal{O}_l$ are odd
according to eq.~(\ref{parities}) and their vacuum expectation values vanish in the continuum
(because of chiral symmetry). We conclude that in this case 
$\langle \mathcal{O} \rangle |_{r,m_{\rm q}}$ is free from O($a$) effects in the chiral limit.
Conversely, if $P_{R_5}^{\mathcal{O}}= 1 \; {\rm mod}\; (2)$ the continuum
limit of its vacuum expectation value (vanishing for symmetry reasons) is approached
with a rate proportional to $a$.
\item Chiral symmetry is realized \`a la Goldstone.
In this case, due to the non-analyticity at $\mq=0$, the chiral point
can only be approached through a limiting procedure. Still,
``automatically'' O($a$) improved correlation functions can be obtained using 
Wilson- or mass-averages or, more practically,  by employing twisted mass fermions at
maximal twist~\cite{Frezzotti:2003ni,AoBa,FPRM}.
\end{itemize}
All these considerations apply to any fermionic theory regularized \`a la Wilson.
In particular we want to numerically test the first scenario described above by considering 
the Schwinger model~\cite{Schw} with two dynamical 
flavors, such that the $R_5$ transformation is well defined.
More importantly, in two dimensions continuous chiral symmetry cannot be spontaneously broken
due to the Mermin-Wagner theorem~\cite{Mermin:1966fe}.
Unfortunately, mainly for numerical reasons, we will not be able to work 
with massless fermions. Therefore in addition to the O($a^2$) cutoff effects expected in 
the chiral limit we might observe O($a\mq$) effects on our quantities.

\vspace{0.22cm}
\ssect{2}
We simulated two dimensional compact QED on a torus with periodic boundary
conditions in time and space. Since the gauge coupling $g$ is of
mass dimension one, the model is super-renormalizable. For the lattice
theory this implies that the continuum limit in a finite physical
volume can be taken at fixed $g\cdot L$ (see
ref.~\cite{Knechtli:2003yt}). For later usage we introduce the
dimensionless coupling $\beta=(ag)^{-2}$.
Clearly, when taking the continuum limit a suitable fermion mass $m$ has to be kept
fixed as well. We decided to define $m$ through the PCAC
relation~\cite{Luscher:1996sc,WISchwing,Gattringer:1999gt} (see also below for
details) and fixed the  product $m \cdot L$ to a constant value.
Notice that due to the super-renormalizability of the model we do not
need to compute renormalization factors $Z$ and can just use the bare
PCAC mass. Indeed, in perturbation theory $Z$
can be written as $Z=1+Z^{(1)} a^2g^2 + \dots$, therefore, at fixed
$g$, loop corrections only change the O($a^2$) ambiguities.
Similarly, if we wanted to fully O($a$) improve the theory (and remove
O($a$) effects also from  vacuum
expectation values of $R_5$ odd operators) by adding
the Sheikholeslami-Wohlert term~\cite{Sheikholeslami:1985ij} to the
action, its coefficient could be set to 1 to all orders in the
perturbative expansion. The same is true for the O($a$)
counterterms of the operators. In other words the O($a$) cutoff effects, if
any, are tree level cutoff effects.

For the simulations we used the Hybrid Monte Carlo
algorithm~\cite{Duane:1987de} with a leapfrog integration scheme.
Observables are constructed from the correlation functions (we set
$a=1$ and write $x=(x_0,{\bf x})$)
\be
C_{\rm XY}(x_0)= {{1}\over{12L^3}} \sum \limits_{\bf x,y,z} 
\psibar(0,{\bf x}) \prod \limits_{i={\bf x}}^{{\bf y}-1} U_1(0,i)
\Gamma_{\rm X} \tau^a \psi(0,{\bf y})\; \psibar(x_0,{\bf z})
\Gamma_{\rm Y} \tau^a \psi(x_0,{\bf z})\;, 
\label{corr}
\ee
with X and Y = A or P and
\be
\Gamma_{\rm A}= \gamma_0 \gamma_5\;, \qquad \Gamma_{\rm P} =  \gamma_5
\;,
\ee
while the Pauli matrices $\tau^a$ act on flavor indices.
In eq.~(\ref{corr}) the product of space-like gauge links is needed to
define gauge invariant wall-to-wall correlators. The additional numerical effort
required to construct such correlation functions is quite moderate in two dimensions.

The PCAC mass $m$ is computed through the ratio  
\be
{{\partial_0 C_{\rm PA}(x_0)}\over{2  C_{\rm PP}(x_0)}}=m \; ,
\ee
derived from the axial Ward identity. Our scaling quantities are obtained from the
correlator $C_{\rm AA}(x_0)$, which for $x_0$ around $T/2$ is expected to be dominated by the 
lowest zero momentum  state $\pi$ in the pseudoscalar sector. In this case, 
the correlator is described by 
\be
C_{\rm AA}(x_0)=\Phi_{\pi}^2 \cosh(m_{\pi}(T/2-x_0))\;, \quad {\rm for} \;\;x_0\simeq T/2 \;,
\label{coshf}
\ee
where the $\cosh$ function is due to the periodicity in time and the matrix element 
$\Phi_{\pi}$ is, up to the normalization,  the analogon  of the pion decay constant $F_{\pi}$ in QCD. 
We will see in the next section
that the formula in eq.~(\ref{coshf}) reproduces the data fairly well, which is plausible 
as we have $m_{\pi}L\simeq 5$ and $T=2L$. 
Since the correlator $C_{\rm AA}(x_0)$ is clearly even under $R_5$
we expect the dimensionless quantities $Lm_{\pi}$ and $L\Phi_{\pi}$ to approach their continuum 
limit values with a rate proportional to $a^2$ up to corrections of O($a\mq$).

\vspace{.22cm}
\ssect{3}
The simulation parameters are collected in table~\ref{sim_par} together with the results.
We also give some details concerning the algorithm.
The hopping parameter 
$\kappa=(2m_0+4)^{-1}$ is tuned in order the keep the fermion mass constant within 
 1\% accuracy.
\begin{table}[htb]
\centering
\begin{tabular}{c l c l l l c c c}
\hline\\[-2ex]
$L/a$      & $\beta$   &$\kappa$ &$m  L$     & $m_{\pi} L$   &$ \Phi_\pi  L $ & $n_{\rm step}$
&  $\#$traj. & accept.  \\
\hline\\[-1.5ex]
16  & 2     &0.2680   & 1.01(1) & 4.80(6)  &0.0400(7) & 50  & 5000  & 96\% \\[0.3ex]
20  & 3.125 &0.2603   & 1.00(1) & 4.7(1)   &0.035(1)  & 50  & 5000  & 94\% \\[0.3ex]
24  & 4.5   &0.2564   & 1.008(7)& 4.7(1)   &0.0321(6) & 50  & 4000  & 93\% \\[0.3ex]
32  & 8     &0.2530   & 0.995(6)& 4.7(1)   &0.0276(9) & 60  & 2500  & 93\% \\[0.3ex]
40  &12.5   &0.25153  & 1.004(8)& 4.68(8)  &0.0269(9) & 70  & 1500  & 92\% \\[0.3ex]
\hline\\[-2ex]
\end{tabular}
\caption{Simulation parameters and results. The length of the single trajectory is always 1, 
discretized in $n_{\rm step}$ intervals.}
\label{sim_par}
\end{table}
To extract the pseudoscalar mass we define a local effective mass, which assumes the correlator
$C_{\rm AA}(x_0)$ to be dominated by a single state, and we average it over a plateau region.
Similarly, we compute $\Phi_{\pi}$ by averaging the ratio $[C_{\rm AA}(x_0)/\cosh(m_\pi (T/2-x_0))]^{1/2}$
over the same region. 
\begin{figure}[htb]
\hspace{0cm}
\vspace{-0.0cm}
\centerline{
\epsfig{file=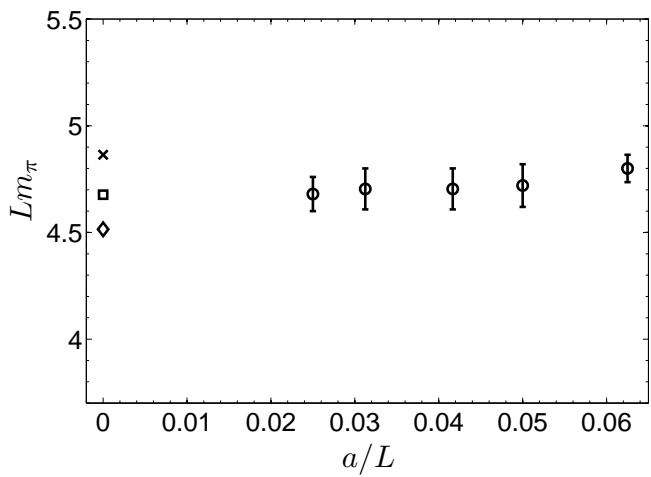,width=10cm,angle=0}
}
\vspace{-1.8cm}
\caption{Scaling plot for the pseudoscalar mass $m_\pi$.
\label{f_pi}}
\end{figure}
%
The results for $Lm_\pi$ are plotted in figure~\ref{f_pi} against $a/L$. It is clear from the plot
that within the 2\% errors we do not see any cutoff effect. The symbols at $a/L=0$ correspond to
the predictions obtained for our value of the fermion mass from different approximate analytical
 solutions valid in the limit of small mass and large coupling $g$~\cite{Smilga}. We regard the 
observed consistency  as a check of our setup. 
In addition, for the same quantity and for a similar choice of parameters, results consistent
with lattice artifacts linear in $a^2$ have been recently reported
also in ref.~\cite{Christian:2005yp}.
%
\begin{figure}[htb]
\hspace{0cm}
\vspace{-0.0cm}
\centerline{
\epsfig{file=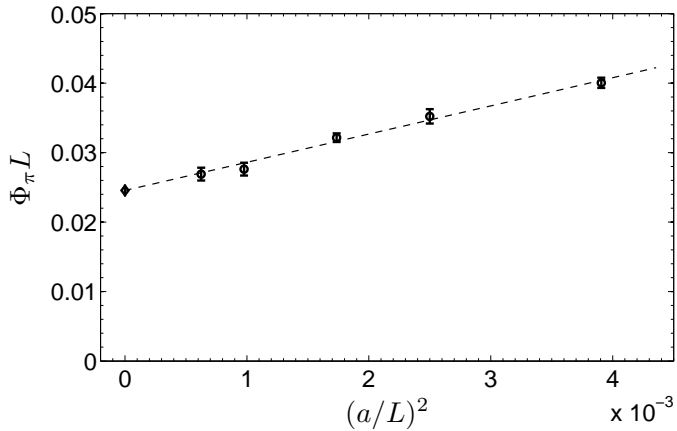,width=10cm,angle=0}
}
\vspace{-1.1cm}
\caption{Scaling plot for the matrix element $\Phi_\pi$.
\label{f_phi}}
\end{figure}
%

The discussion of $L\Phi_\pi$ is a bit more delicate since we see cutoff effects in 
this quantity. As it is shown in figure~\ref{f_phi}, those are clearly consistent with being 
linear in $a^2$ only. Nevertheless, in order to estimate the size of the O$(a\mq)$ effects,
we tried to fit the data also to a polynomial with terms linear and quadratic in $a$. 
The fit is acceptable in terms of $\chi^2$ and the continuum limit we obtain is in
agreement with the one in figure~\ref{f_phi}, but it has a five times
larger error. The coefficients of the linear and quadratic terms have
large errors as well. They are both consistent with zero but
strongly anticorrelated. We conclude that the sensitivity
of our data to the O($a\mq$) effects is very small.
Adding a smaller lattice
resolution would probably help to disentangle them from the O($a^2$).

\vspace{0.22cm}
\ssect{4}
The numerical study presented here confirms the expectation that in the absence of spontaneous
 chiral symmetry breaking cutoff effects are of O($a^2$) also when Wilson fermions are used
(at least if the chiral limit is considered).
The situation is very different from QCD in four dimensions (and large volume), where for
Wilson fermions the O($a$) effects are rather large~\cite{advLQCD} and have to be removed by following the 
Symanzik improvement programme~\cite{Luscher:1996sc}.

As a consequence,  testing fermionic actions by scaling studies in the Schwinger model 
provides, in our opinion, very little information about the cutoff effects for the same 
regularizations in the phenomenologically more relevant case of QCD.

On the other hand, to improve our confidence in the argument presented here,
it would be interesting to extend the study by considering different values 
of the fermion mass in order to assess more precisely the size of the residual O($a\mq$)
effects. As far as we can tell now, those appear to be fairly small.
In addition, the mass of the scalar particle $\eta$ could be included among the observables.
Contrary to the quantities discussed here, this mass does not vanish
in the chiral limit~\cite{Gattringer:1999gt}.
To this end, the numerical techniques introduced in ref.~\cite{Foley:2005ac} could provide
an efficient way to evaluate the contributions coming from disconnected quark diagrams.

\vspace{0.2cm}
\noindent{\bf{Acknowledgments}}. We thank Francesco Knechtli, Tomasz Korzec, Rainer Sommer and
Ulli Wolff for useful discussions. We thank Ulli Wolff for a critical
reading of the manuscript. The simulations in the present work have been carried out using
the PC cluster at the Humboldt University in Berlin, we thank the staff at the computer center
for the assistance. M.D.M. gratefully acknowledges the SFB Transregio 9 for financial 
support.

\end{document}